\documentstyle[twocolumn,prd,aps]{revtex}
\input{epsf}
\begin{document}

\title{Model-independent search for the Abelian $Z'$
 boson and the LEP data}
\author{A. V. Gulov
 \thanks{Email: gulov@ff.dsu.dp.ua} and
 V. V. Skalozub
 \thanks{Email: skalozub@ff.dsu.dp.ua}}
\address{Dniepropetrovsk National University, Dniepropetrovsk,
 49050 Ukraine}
\date{\today}
\maketitle
\begin{abstract}
Model-independent search for the Abelian $Z'$ gauge boson virtual
state based on the new observable in the processes $e^+ e^- \to
l^+ l^-$ is analyzed for the LEP data. The observable is
introduced by using following from renormalizability of an
underlying theory the correlations between the parameters of the
low energy effective Lagrangian and kinematics of the scattering
processes. The central values of the observable are in accordance
with the $Z'$ existence, although no real signal is found at the
$1\sigma$ level. Lower limits at the 95\% confidence level on the
corresponding contact interaction scale are derived in the range
of 15.6 - 18.6 TeV. Comparisons with other fits are done.
\end{abstract}

\section{INTRODUCTION}

Among various objectives of the recently finished LEP experiments
an important place was devoted to searching for signals of new
physics beyond the energy scale of the standard model (SM).
Reports on these results are adduced partly in the literature
\cite{osaka}. In the present note we are going to discuss the
problem of searching for the heavy $Z'$ gauge boson \cite{leike}.
This particle is a necessary element of the different models
extending the SM. Low limits on its mass following from the
analysis of variety of popular models ($\chi$, $\psi$, $\eta$,
L--R models \cite{models} and the Sequential Standard Model (SSM)
\cite{SSM}) are found to be in the energy intervals 500--2000 GeV
\cite{osaka} (see Table \ref{t1} which reproduces Table 9 of Ref.
\cite{osaka}). As it is seen, the values of $m_{Z'}$ (as well as
the parameters of interactions with the SM particles) are strongly
model dependent. Therefore, it seems reasonable to find some model
independent signals of this particle. To elaborate that general
principles of field theory must be taken into consideration giving
a possibility to relate the parameters of different scattering
processes. Then, one is able to introduce variables, convenient
for the model independent search for $Z'$ (or other heavy states).
These ideas were used in Ref. \cite{ZprTHDM} in order to introduce
the model-independent sign definite variables for $Z'$ detection
in scattering processes with $\sqrt{s}\simeq 500$ GeV.

As it has been pointed out in Ref. \cite{ZprTHDM}, some parameters
of new heavy fields can be related by using the requirement of
renormalizability of the underlying model remaining in other
respects unspecified. The relations between the parameters of new
physics due to the renormalizability were called the
renormalization group (RG) relations. In Ref. \cite{ZprTHDM} the
RG relations for the low-energy $Z'$ couplings to the SM fields
have been derived. They predict two possible types of $Z'$
particles, namely, the chiral and the Abelian $Z'$ ones. Each type
is described by a few couplings to the SM fields. Therefore,
taking into account the RG correlations between the $Z'$
couplings, one is able to introduce observables which uniquely
pick up the $Z'$ virtual state \cite{ZprTHDM}. In the present
paper we discuss these observables and the constraints on possible
$Z'$ signals following from the analysis of the LEP data.

\section{$Z'$ COUPLINGS TO FERMIONS}

To consider the $Z'$ interactions with light particles one must
specify the model describing physics at low energies. For
instance, the minimal SM with the one scalar doublet can be
chosen. However, due to the lack of information about scalar
fields, models with an extended set of light scalar particles can
be also considered. Below we choose the two-Higgs-doublet model
(THDM) \cite{THDM} as the low-energy theory (notice, the minimal
SM is the particular case of the THDM).

To derive the RG relations one has to introduce the
parametrization of $Z'$ couplings to the SM fields. Since we are
going to account of the $Z'$ effects in the low-energy
$e^+e^-\to\bar{f}f$ processes in lower order in $m^{-2}_{Z'}$, the
linear in $Z'$ interactions with the SM fields are of interest,
only. The renormalizability of the underlying theory and the
decoupling theorem \cite{decoupling} guarantee the dominance of
renormalizable $Z'$ interactions at low energies. The interactions
of the non-renormalizable types, generated at high energies due to
radiation corrections, are suppressed by the inverse heavy mass.
The SM gauge group $SU(2)_L\times U(1)_Y$ is considered as a
subgroup of the underlying theory group. So, the mixing
interactions of the types $Z'W^+W^-$, $Z'ZZ$, ... are absent at
the tree level. Such conditions are usually used in the literature
\cite{EL1} and lead to the following parametrization of the linear
in $Z'$ low-energy vertices:
\begin{eqnarray}\label{1}
 {\cal L}&=&
 \sum\limits_{i=1}^2
  \left|\left(
  D^{{\rm ew,} \phi}_\mu -
  \frac{i\tilde{g}}{2}\tilde{Y}(\phi_i)\tilde{B}_\mu
  \right)\phi_i\right|^2
 +
 \nonumber\\&&
 i\sum\limits_{f=f_L,f_R}\bar{f}{\gamma^\mu}
  \left(
  D^{{\rm ew,} f}_\mu -
  \frac{i\tilde{g}}{2}\tilde{Y}(f)\tilde{B}_\mu
  \right)f,
\end{eqnarray}
where $\phi_i$ $(i=1,2)$ are the scalar doublets, $\tilde{g}$ is
the charge corresponding to the $Z'$ gauge group, $D^{{\rm
ew,}\phi}_\mu$ and $D^{{\rm ew,}f}_\mu$ are the electroweak
covariant derivatives, $\tilde{B}_\mu$ denotes the massive $Z'$
field before the spontaneous breaking of the electroweak symmetry,
and the summation over the all SM left-handed fermion doublets,
$f_L =\{(f_u)_L, (f_d)_L\}$, and the right-handed singlets, $f_R =
(f_u)_R, (f_d)_R$, is understood. Diagonal $2\times 2$ matrices
$\tilde{Y}(\phi_i)$, $\tilde{Y}(f_L)$ and numbers $\tilde{Y}(f_R)$
are unknown $Z'$ generators characterizing the model beyond the
SM.

The one-loop RG relations for the above introduced $Z'$ vertices
(\ref{1}) have been obtained in Ref. \cite{ZprTHDM}. As it was
shown, two different types of the $Z'$ generators are compatible
with the renormalizability of the underlying theory. The first
type, called the chiral $Z'$, describes the $Z'$ boson which
couples to the SM doublets, only. The corresponding generators
have the zero traces:
\begin{eqnarray}\label{chiral}
 &&\tilde{Y}(\phi_i)=
 -\tilde{Y}_{\phi}
 \left(\begin{array}{cc}
 1 & 0 \\ 0 & -1
 \end{array}\right),
 \nonumber\\&&
 \tilde{Y}(f_L)=
 \tilde{Y}_{L,f_u}
 \left(\begin{array}{cc}
 1 & 0 \\ 0 & -1
 \end{array}\right),
 \nonumber\\&&
 \tilde{Y}(f_R)=0.
\end{eqnarray}
The second type is the Abelian $Z'$ boson:
\begin{eqnarray}\label{abelian}
 &&\tilde{Y}(\phi_i)=
 \tilde{Y}_{\phi}
 \left(\begin{array}{cc}
 1 & 0 \\ 0 & 1
 \end{array}\right),
 \nonumber\\&&
 \tilde{Y}(f_L)=
 \tilde{Y}_{L,f}
 \left(\begin{array}{cc}
 1 & 0 \\ 0 & 1
 \end{array}\right),
 \nonumber\\&&
 \tilde{Y}(f_R)=
 \tilde{Y}_{L,f}+2T^3_f \tilde{Y}_{\phi},
\end{eqnarray}
where $T^3_f$ is the third component of the fermion weak isospin.
The relations (\ref{abelian}) ensure, in particular, the
invariance of the Yukawa terms with respect to the effective
low-energy $\tilde{U}(1)$ subgroup corresponding to the $Z'$
boson. As it follows from the relations, the couplings of the
Abelian $Z'$ to the axial-vector fermion currents have the
universal absolute value proportional to the $Z'$ coupling to the
scalar doublets.

The derived relations (\ref{chiral})--(\ref{abelian}) are
independent of a specific model beyond the SM predicting the $Z'$
boson. They hold in the THDM as well as in the minimal SM. As it
is seen from relations (\ref{chiral})--(\ref{abelian}), only one
parameter for each SM doublet remains arbitrary. The rest
parameters are expressed through them. A few number of independent
$Z'$ couplings gives the possibility to introduce the observables
convenient for detecting uniquely the $Z'$ signals in experiments
\cite{ZprTHDM,obs}. In what follows, we analyze the obtained at
LEP data taking into account the RG relations (\ref{abelian}) in
order to constrain possible signals of the Abelian $Z'$ boson.

\section{OBSERVABLES FOR THE ABELIAN $Z'$ SEARCH}

Consider the leptonic processes $e^+e^-\to V^\ast\to l^+l^-$
($l=\mu,\tau$) with the neutral vector boson exchange
($V=A,Z,Z'$). We assume the non-polarized initial- and final-state
fermions. At LEP energies $\sqrt{s}\simeq 200$ GeV the fermions
can be treated as massless particles, $m_e,m_l\sim 0$. In this
approximation the left-handed and the right-handed fermions can be
substituted by the helicity states, which we mark as $\lambda$ and
$\xi$ for the incoming electron and the outgoing fermion,
respectively ($\lambda,\xi=L,R$).

The differential cross section of the process $e^+e^-\to V^\ast\to
l^+l^-$ deviates from its SM value by a quantity of order
$m^{-2}_{Z'}$:
\begin{eqnarray}
 \Delta\frac{d\sigma_l}{d\cos\theta}&=&
 \frac{1}{16\pi s} \mbox{Re}\Big[
 {\cal A}^\ast_{\rm SM}
 \nonumber\\&&\times
 \left({\cal A}_{Z^\prime} +
 \left.\frac{d{\cal A}_Z}{d\theta_0}\right|_{\theta_0 =0}
 \theta_0 \right)\Big],
 \nonumber\\
 {\cal A}_{\rm SM}&=&{\cal A}_A + {\cal A}_Z(\theta_0 =0),
\end{eqnarray}
where $\theta$ denotes the angle between the momentum of the
incoming electron and the momentum of the outgoing lepton, ${\cal
A}_V$ is the Born amplitude of the process, and $\theta_0\sim
m^2_W/m^2_{Z'}$ is the $Z$--$Z'$ mixing angle. The leading
contribution comes from the interference between the $Z'$ exchange
amplitude, ${\cal A}_{Z'}$, and the SM amplitude, ${\cal A}_{\rm
SM}$, whereas the $Z$--$Z'$ mixing terms are suppressed by the
additional small factor $m^2_Z/s$. Notice that the deviation
$\Delta d\sigma_l/d\cos\theta$ depends on the center-of-mass
energy through the quantity $m^2_Z/s$, only.

To take into consideration the correlations (\ref{abelian}) let us
introduce the observable $\sigma_l(z)$ defined as the difference
of cross sections integrated in some ranges of the scattering
angle $\theta$, which will be specified below \cite{ZprTHDM,obs}:
\begin{eqnarray}\label{eq8}
 \sigma_l(z)
 &\equiv&\int\nolimits_z^1
  \frac{d\sigma_l}{d\cos\theta}d\cos\theta
 -\int\nolimits_{-1}^z
  \frac{d\sigma_l}{d\cos\theta}d\cos\theta
 \nonumber\\&=&
 \sigma^T_l\left[ A^{FB}_l\left(1-z^2\right)
 -\frac{z}{4}\left(3+z^2\right)\right],
\end{eqnarray}
where $z$ stands for the boundary angle, $\sigma^T_l$ denotes the
total cross section and $A^{FB}_l$ is the forward-backward
asymmetry of the process. The idea of introducing the
$z$-dependent observable (\ref{eq8}) is to choose the value of the
kinematic parameter $z$ in such a way that to pick up the
characteristic features of the Abelian $Z'$ signals. Due to the
correlations between the Abelian $Z'$ couplings the quantity
(\ref{eq8}) can be written as follows
\begin{eqnarray}\label{obs2}
 \Delta\sigma_l(z)
 &=& \frac{\alpha_{\rm em}\tilde{g}^2}{16m^2_{Z'}}
 \left(
 {\cal F}^l_0(z,s)a^2_{Z'}
 \right.\nonumber\\&&
 +{\cal F}^l_1(z,s)v^l_{Z'}v^e_{Z'}
 +{\cal F}^l_2(z,s)v^l_{Z'}|a_{Z'}|
 \nonumber\\&&\left.
 +{\cal F}^l_3(z,s)v^e_{Z'}|a_{Z'}|
 \right).
\end{eqnarray}
where $v^l_{Z'}\equiv(\tilde{Y}_{L,l}+\tilde{Y}_{R,l})/2$ and
$a^l_{Z'}\equiv (\tilde{Y}_{R,l}-\tilde{Y}_{L,l})/2=
T^3_f\tilde{Y}_{\phi}$ are the $Z'$ couplings to the vector and
the axial-vector lepton currents. Functions ${\cal F}^l_i(z,s)$
are determined by the SM quantities and independent of the lepton
generation. The leading contributions to the factors ${\cal
F}^l_2(z,s)={\cal F}^l_3(z,s)$ equal to zero. So, by choosing the
boundary angle $z^\ast$ to be the solution to the equation ${\cal
F}^l_1(z^\ast,s)=0$, one can switch off three factors ${\cal
F}^l_i(z,s)$ $(i=1,2,3)$ simultaneously. The function $z^\ast(s)$
is the decreasing function of energy. This is shown in Table
\ref{t2} for the LEP energies. In Table \ref{t2} we also show the
factors ${\cal F}^l_i(z^\ast,s)$. As it is occurred, these factors
contribute less than 2\%.

By choosing $z=z^\ast$, we obtain the sign definite observable
\begin{eqnarray}\label{eq7}
 \Delta\sigma_l(z^\ast)&\simeq&
 \frac{\alpha_{\rm em}\tilde{g}^2}{16 m^2_{Z^\prime}}
 {\cal F}^l_0(z^\ast,s)a^2_{Z^\prime}<0.
\end{eqnarray}
The quantity $\Delta\sigma_l(z^\ast)$ is negative and the same for
the all types of the SM charged leptons. This is the
model-independent signal of the Abelian $Z'$ boson. Thus, the
variables $\Delta\sigma_l(z^\ast)$ select the $Z'$ boson signals
in the processes $e^+e^-\to l^+l^-$.

\section{ANALYSIS OF THE LEP DATA}

The measurements of the cross-sections and the forward-backward
asymmetries have been combined for the full LEP2 data set
recently. The preliminary results are adduced in
Ref.~\cite{osaka}. Let us analyze these data assuming a model
independent search for signals of the Abelian $Z'$ boson.

Before doing that in the way outlined in the previous sections, we
note that in Ref. \cite{osaka} some other analysis of signals of
physics beyond the SM in the processes $e^+e^-\to\bar{f}f$ was
present. It is based on the introduction of the ``models''
describing different kinds of new four-fermion contact
interactions. Since one is able to successfully fit only one
parameter of new physics, eight models (LL, RR, LR, RL, VV, AA,
A0, V0) assuming various helicity coupling between the initial
state and final state currents are discussed. Each model is
described by only one non-zero coupling. For example, in the LL
model the non-zero coupling of left-handed fermions is taken into
account. The signal of a new heavy particle is fitted by
considering the interference of the SM amplitude with the contact
four-fermion term. Whatever physics beyond the SM exists, it can
manifest itself in some contact coupling mentioned. Hence, it is
possible to find a low limit on the masses of the states
responsible for the interactions considered. In principle, a
number of states may contribute into each of the models.
Therefore, the purpose of the fit described by these models is to
find any signal of new physics. No specific types of new particles
are considered in this analysis. The virtual states of heavy
particles (for instance, the Abelian $Z'$ boson) contribute to
several contact interactions simultaneously, and the corresponding
couplings cannot be switched off separately.

It is interesting to note the fits for the process
$e^+e^-\to\mu^+\mu^-$ in Ref. \cite{osaka}. Several models
mentioned demonstrate one standard deviation from the SM
predictions for this process. In this regard, we note the paper
\cite{bourilkov} in which the mentioned models were investigated
for the Bhabha scattering $e^+e^-\to e^+e^-(\gamma)$. The
deviations from the SM at the $1\sigma$--level were found again,
whereas the AA model shows the $2\sigma$--level deviation.
However, the question whether these deviations could be interpret
as a signal of the Abelian $Z'$ remains open.

The aim of our investigation is to pick up a model-independent
signal of the virtual Abelian $Z'$ boson, which uniquely selects
this particle. The strategy is similar - to introduce the proper
one-parametric variable (\ref{eq7}) on the discussed principles of
renormalizability and the kinematic properties of scattering
amplitudes. In our case we have a possibility to derive this
observable by taking into consideration the renormalizability
relations (\ref{abelian}) and, therefore, the selection of the
axial-vector currents is not the ``model''. The observable $\Delta
\sigma_l (z^\ast)$ accounts for all interactions of the $Z'$
boson. Below, we will make an additional comparison of our method
and that of in Ref. \cite{osaka}.

Now, to search for the model-independent $Z'$ signals we will
analyze the observable $\Delta\sigma_l (z^\ast)$ on the base of
the full LEP data set. It depends on one unknown $Z'$ parameter,
$\tilde{g}^2 a^2_{Z'}/m^2_{Z'}$, which describes the effective
contact interaction between the axial-vector lepton currents. In
what follows we will use the notation
\begin{equation}\label{e8}
\frac{\tilde{g}^2 a^2_{Z'}}{16\pi m^2_{Z'}}\equiv
\frac{1}{\Lambda^2}\equiv\epsilon.
\end{equation}
This normalization is admitted, in particular, in Ref.
\cite{osaka}. The parameter $\epsilon$ is related to the
observable $\Delta\sigma_l(z^\ast)$ by the factor which slightly
depends on energy (see Table \ref{t2}):
\begin{equation}
\epsilon=\frac{\Delta\sigma_l(z^\ast)}
 {\pi\alpha_{\rm em}{\cal F}^l_0(z^\ast,s)}.
\end{equation}

Treating the observable $\Delta\sigma_l(z^\ast)$ has the following
advantages:
\begin{enumerate}
\item
All the LEP data for the processes $e^+e^-\to l^+l^-$ can be
incorporated to obtain the limits on the observable.
\item
There is one parameter of new physics to be fitted.
\item
The data for the $e^+e^-\to\mu^+\mu^-$ and $e^+e^-\to\tau^+\tau^-$
scattering can be used to measure the same observable.
\item
The sign of the observable ($\epsilon >0$) is the characteristic
feature of the Abelian $Z'$ signal.
\end{enumerate}

\begin{figure}
\epsfxsize=0.35\textwidth \epsfbox[0 0 600 600]{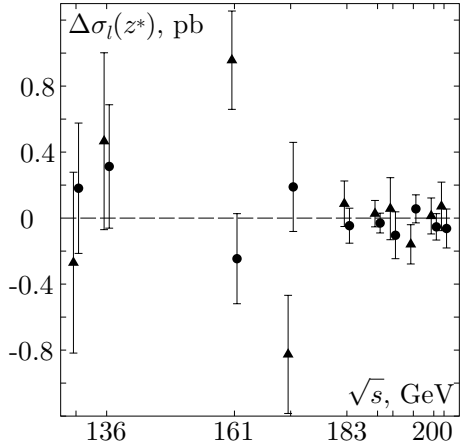}
\caption{$\Delta\sigma_l(z^\ast)$ computed from the LEP data. The
circles and triangles represent the measurements for the
$e^+e^-\to\mu^+\mu^-$ and $e^+e^-\to\tau^+\tau^-$ processes,
respectively.} \label{fig:zast}
\end{figure}

The LEP data for the total cross-sections and the forward-backward
asymmetries as well as the SM values of these quantities are shown
in Tables \ref{t3}--\ref{t4}. From the set of the data we compute
the observable $\Delta\sigma_l(z^\ast)$ and the corresponding
error $\delta\sigma_l(z^\ast)$ for each LEP energy by means of the
following relations
\begin{eqnarray}
 \Delta\sigma_l(z^\ast)
 &=&
 \left[
 A_l^{FB}\left(1-z^{\ast 2}\right)
 -\frac{z^\ast}{4}\left(3 +z^{\ast 2}\right)
 \right] \Delta\sigma_l^T
 \nonumber\\&&
 + \left(1 - z^{\ast 2}\right)
  \sigma_{l,\rm SM}^T \Delta A_l^{FB},
 \nonumber\\
 \delta\sigma_l(z^\ast)^2
 &=&
 {\left[
 A_l^{FB}\left(1-z^{\ast 2}\right)
 -\frac{z^\ast}{4}\left(3 +z^{\ast 2}\right)
 \right]}^2 (\delta\sigma_l^T)^2
 \nonumber\\&&
 +{\left[
 \left(1 - z^{\ast 2}\right)
 \sigma_{l,\rm SM}^T
 \right]}^2 (\delta A_l^{FB})^2.
\end{eqnarray}
The results are given in Table \ref{t5} and Fig. \ref{fig:zast}.
All the values of the observable are no more than one standard
deviation from the SM value except for two points at 161 and 172
GeV corresponding to the $e^+e^-\to\tau^+\tau^-$ process. These
points reflect the significant dispersion of the measurements for
the scattering into $\tau$ pairs at $\sqrt{s}< 183$ GeV. As it is
also seen from Fig. \ref{fig:zast}, the measurements for the
center-of-mass energies $\sqrt{s}\ge 183$ GeV have a higher level
of precision.

We will choose three different sets of data to fit the parameter
$\epsilon$. The first one is the complete set containing 20
points. The second set includes the data for the center-of-mass
energies $\sqrt{s}\ge 183$ GeV (12 measurements). The final set
contains the data for the scattering into $\mu$ pairs (10 points).

Thus, there are 20 measurements $\epsilon_i$ ($i=1,\ldots,20$) of
the parameter $\epsilon$ with unequal precisions. We define the
fitted value of $\epsilon$, $\bar\epsilon$, as a linear
combination of $\epsilon_i$ with factors which minimize the
corresponding dispersion, $(\delta\bar\epsilon)^2$. It is easy to
show that
\begin{equation}
 \bar\epsilon=
 {\left[
  \sum_i \frac{1}{(\delta\epsilon_i)^2}
  \right]}^{-1}
 \sum_i \frac{\epsilon_i}{(\delta\epsilon_i)^2},
\end{equation}
\begin{equation}
 (\delta\bar\epsilon)^2= {\left[
  \sum_i \frac{1}{(\delta\epsilon_i)^2}
  \right]}^{-1}.
\end{equation}
The value of $\delta\bar\epsilon$ gives the $1\sigma$-level
interval for $\bar\epsilon$. Note that the same value of
$\bar\epsilon$ can be derived as a result of the minimization of
the likelihood function
\begin{equation}
-\log{\cal L}(\epsilon)=
\sum_i\frac{(\epsilon_i-\epsilon)^2}{2(\delta\epsilon_i)^2},
\end{equation}
or the corresponding $\chi^2$-function. In this case the
$1\sigma$-level interval ($\bar\epsilon-\delta\bar\epsilon$,
$\bar\epsilon+\delta\bar\epsilon$) could be alternatively derived
from the condition $\log[{\cal L}(\epsilon)/{\cal
L}(\bar\epsilon)]=-0.5$.

We use the log-likelihood method to determine a one sided lower
limit on the scale $\Lambda$ at the 95\% confidence level. It is
derived by the integration of the likelihood function over the
physically allowed region $\epsilon>0$. The exact definition is
\begin{equation}
\int_{0}^{1/\Lambda^2}{\cal L}(\epsilon ')d\epsilon ' =
0.95\int_{0}^{\infty}{\cal L}(\epsilon ')d\epsilon '.
\end{equation}

In Table \ref{t6} we show the fitted values of $\epsilon$ with
their 68\% confidence level uncertainties and the 95\% confidence
level lower limit on the scale $\Lambda$. We also compute the
total probability of the $Z'$ signal by integrating the likelihood
function over the whole allowed range of $\epsilon$
($\epsilon>0$).

As it is seen, all data sets lead to the comparable fitted values
of $\bar\epsilon$ with the nearly equal uncertainties. All the
central values $\bar\epsilon$ have the sign compatible with the
$Z'$ signal. However, they are no more than one standard deviation
from the SM value $\bar\epsilon=0$. Thus, though the central
values witness to the $Z'$ existence, no $Z'$ signal is detected
at the $1\sigma$ confidence level. This result reflects the lack
of the accuracy of the input data. In this regard, fitting the
observable directly from the final differential cross sections
when they will be published may probably improve the accuracy.

The more precise data corresponding to the scattering into
$\mu^+\mu^-$ pairs demonstrate the largest positive central value
of $\epsilon$. This value is nearly one standard deviation. Taking
into account the data for $\tau^+\tau^-$ final states decreases
the central value of the $Z'$ signal but does not affect
essentially the uncertainty of the results.

As our investigation showed, the characteristic signal of the
Abelian $Z'$ boson is concerned with the coupling of axial-vector
currents. In this regard, let us turn again to the helicity
``models'' of Ref. \cite{osaka} and compare our results with the
fitting for the AA case. As it follows from the present analysis
this model is sensitive mostly to the signals of the Abelian $Z'$
boson. Of course, the parameter $\epsilon$ in Ref. \cite{osaka}
and in Eq. (\ref{eq8}) is not the same quantity. As we already
noted, in the AA model the $Z'$ couplings to the vector fermion
currents are set to zero, therefore it is able to describe only
some particular case of the Abelian $Z'$ boson. Moreover, in this
model both the positive and the negative values of $\epsilon$ are
considered, whereas in our approach only the positive ones are
permissible. As the value of four-fermion contact coupling in the
AA model is dependent on the lepton flavor, the Abelian $Z'$
induces the axial-vector coupling universal for all lepton types.
Nevertheless, it is interesting to note that the fitted value of
$\epsilon$ in the AA model for the $\mu^+\mu^-$ final states
($-0.0033^{+0.0032}_{-0.0012}\mbox{ TeV}^{-2}$) as well as the
value derived under the assumption of the lepton universality
($-0.0013^{+0.0018}_{-0.0015}\mbox{ TeV}^{-2}$) are similar to our
results. Since the parameters $\epsilon$ in Ref. \cite{osaka} and
in Eq. (\ref{e8}) have opposite signs by definition, the signs of
the central values in the AA model agree with those of Table
\ref{t6}, whereas the uncertainties are of the same order. Thus,
as it follows from this analysis, the AA model is mainly
responsible for signals of the Abelian $Z'$ gauge boson although a
lot of details concerning its interactions is not accounted for
within this fit.

The $Z'$ boson mass is related to the contact interaction scale
$\Lambda$ by Eq. (\ref{e8}). To convert the scale $\Lambda$ to the
$Z'$ mass one have to assume the value of coupling $\tilde{g}$.
When the $Z'$ boson couples to the SM particles with a strength
comparable with the electroweak forces $\tilde{g}\simeq g$, the
central values of $\bar\epsilon$ correspond to the masses of order
3--5 TeV, whereas the lower limit on $m_{Z'}$ is about 1.4--1.7
TeV. Now let us compare these values with the constraints for
different models in Table \ref{t1}. As it is seen, the Abelian
$Z'$ cannot be so light as the $\chi$, $\psi$, and $\eta$ models
predict. On the other hand, the lower limits derived in our paper
are close to ones of the L--R models and the Sequential Standard
Model. Thus, although the $Z'$ boson is not detected at LEP, it
can be light enough to be discovered at LHC.

\begin{table}
\centering \caption{95\% confidence level lower limits on the $Z'$
mass for some popular models.}\label{t1}
\begin{tabular}{|c|c|c|c|c|c|}
 Model & $\chi$ & $\psi$ & $\eta$ & L--R & SSM \\ \hline
 $m^{\rm limit}_{Z'}, {\rm GeV}/c^2$ &
 630 & 510 & 400 & 950 & 2260 \\
\end{tabular}\end{table}

\begin{table}
\centering \caption{The boundary angle $z^\ast$ and the factors
${\cal F}^l_i(z^\ast,s)$ computed at energies of the LEP
experiments.} \label{t2}
\begin{tabular}{|l|c c c|}
 $\sqrt{s}$, GeV & $z^\ast$ & ${\cal F}^l_0(z^\ast,s)$ &
 ${\cal F}^l_{2,3}(z^\ast,s)$ \\ \hline
 130 & 0.488 & -2.349 & -0.051 \\
 136 & 0.466 & -2.456 & -0.053 \\
 161 & 0.409 & -2.381 & -0.049 \\
 172 & 0.393 & -2.300 & -0.047 \\
 183 & 0.381 & -2.225 & -0.045 \\
 189 & 0.376 & -2.187 & -0.045 \\
 192 & 0.374 & -2.169 & -0.044 \\
 196 & 0.370 & -2.146 & -0.044 \\
 200 & 0.368 & -2.124 & -0.043 \\
 202 & 0.366 & -2.114 & -0.043 \\
\end{tabular}\end{table}

\begin{table}
\centering \caption{The LEP data for $\sigma^T_l$ (pb) and
$A^{FB}_l$} \label{t3}
\begin{tabular}{|l|c c c c|}
 $\sqrt{s}$, GeV & $\sigma^T_l$ &
 $\delta\sigma^T_l$ & $A^{FB}_l$ & $\delta A^{FB}_l$
 \\ \hline \multicolumn{5}{|c|}{$l=\mu$}\\ \hline
 130 & 8.331 & 0.664 & 0.736 & 0.059 \\
 136 & 8.229 & 0.678 & 0.709 & 0.062 \\
 161 & 4.585 & 0.364 & 0.546 & 0.07 \\
 172 & 3.555 & 0.317 & 0.679 & 0.077 \\
 183 & 3.484 & 0.147 & 0.558 & 0.035 \\
 189 & 3.108 & 0.077 & 0.565 & 0.021 \\
 192 & 2.925 & 0.177 & 0.538 & 0.052 \\
 196 & 2.948 & 0.107 & 0.585 & 0.032 \\
 200 & 3.052 & 0.107 & 0.522 & 0.032 \\
 202 & 2.622 & 0.14 & 0.541 & 0.048 \\
 \hline \multicolumn{5}{|c|}{$l=\tau$}\\ \hline
 130 & 9.065 & 0.927 & 0.652 & 0.084 \\
 136 & 7.123 & 0.821 & 0.771 & 0.088 \\
 161 & 5.692 & 0.545 & 0.769 & 0.063 \\
 172 & 4.026 & 0.45  & 0.344 & 0.107 \\
 183 & 3.398 & 0.174 & 0.609 & 0.045 \\
 189 & 3.14 & 0.1 & 0.584 & 0.028 \\
 192 & 2.86 & 0.219 & 0.608 & 0.068 \\
 196 & 2.994 & 0.141 & 0.498 & 0.046 \\
 200 & 2.966 & 0.137 & 0.553 & 0.043 \\
 202 & 2.8 & 0.186 & 0.583 & 0.059 \\
\end{tabular}\end{table}

\begin{table}
\centering \caption{SM values for $\sigma^T_l$ (pb) and
$A^{FB}_l$} \label{t4}
\begin{tabular}{|l|c c|c c|}
 $\sqrt{s}$, GeV & $\sigma^{T,\rm SM}_\mu$ &
 $\sigma^{T,\rm SM}_\tau$ & $A^{FB,\rm SM}_\mu$ &
 $A^{FB,\rm SM}_\tau$  \\ \hline
 130 & 8.439 & 8.435 & 0.705 & 0.704 \\
 136 & 7.281 & 7.279 & 0.684 & 0.683 \\
 161 & 4.613 & 4.613 & 0.609 & 0.609 \\
 172 & 3.952 & 3.951 & 0.591 & 0.591 \\
 183 & 3.446 & 3.446 & 0.576 & 0.576 \\
 189 & 3.207 & 3.207 & 0.569 & 0.569 \\
 192 & 3.097 & 3.097 & 0.566 & 0.566 \\
 196 & 2.962 & 2.962 & 0.562 & 0.562 \\
 200 & 2.834 & 2.833 & 0.558 & 0.558 \\
 202 & 2.77  & 2.769 & 0.556 & 0.556 \\
\end{tabular}\end{table}

\begin{table}
\centering \caption{$\Delta\sigma_l(z^\ast)$ (pb) and $\epsilon$
(${\rm TeV}^{-2}$) from the LEP data} \label{t5}
\begin{tabular}{|l|c c|c c|}
 $\sqrt{s}$, GeV & $\Delta\sigma_l(z^\ast)$ &
 $\delta\sigma_l(z^\ast)$ & $\epsilon$ & $\delta\epsilon$
 \\ \hline
 \multicolumn{5}{|c|}{$l=\mu$}\\ \hline
 130 & 0.181 & 0.395 & -0.0081 & 0.0176 \\
 136 & 0.313 & 0.374 & -0.0133 & 0.0159 \\
 161 & -0.246 & 0.273 &  0.0108 & 0.0120 \\
 172 & 0.189 & 0.270 & -0.0086 & 0.0123 \\
 183 & -0.046 & 0.106 &  0.0022 & 0.0050 \\
 189 & -0.030 & 0.060 &  0.0014 & 0.0029 \\
 192 & -0.104 & 0.142 &  0.0050 & 0.0068 \\
 196 & 0.056 & 0.085 & -0.0027 & 0.0041 \\
 200 & -0.053 & 0.080 &  0.0026 & 0.0040 \\
 202 & -0.063 & 0.118 & 0.0031 & 0.0058 \\
 \hline \multicolumn{5}{|c|}{$l=\tau$}\\ \hline
 130 & -0.270 & 0.548 &  0.0120 & 0.0244 \\
 136 & 0.466 & 0.536 & -0.0198 & 0.0228 \\
 161 & 0.957 & 0.298 & -0.0421 & 0.0131 \\
 172 & -0.826 & 0.358 &  0.0376 & 0.0163 \\
 183 & 0.087 & 0.138 & -0.0041 & 0.0065 \\
 189 & 0.027 & 0.080 & -0.0013 & 0.0038 \\
 192 & 0.057 & 0.188 & -0.0028 & 0.0091 \\
 196 & -0.159 & 0.119 &  0.0078 & 0.0058 \\
 200 & 0.013 & 0.109 & -0.0006 & 0.0053 \\
 202 & 0.071 & 0.147 & -0.0035 & 0.0073 \\
\end{tabular}\end{table}

\begin{table}
\centering \caption{The fitted values of $\epsilon$ and their 68\%
confidence level uncertainties together with the 95\% confidence
level lower limit on $\Lambda$ and the probability of the $Z'$
signal.} \label{t6}
\begin{tabular}{|l|c c|c|}
 Data set & $\bar\epsilon$, ${\rm TeV}^{-2}$ &
 $\Lambda_{\rm min}$, TeV & $P$, \%
 \\ \hline
 All LEP data & $0.00037\pm 0.00134$ & 18.6 & 60.8 \\
 $\sqrt{s}\ge 183$ GeV & $0.00074\pm 0.00138$ & 17.6 & 70.4 \\
 $e^+e^-\to\mu^+\mu^-$ & $0.00114\pm 0.00168$ & 15.6 & 75.2 \\
\end{tabular}\end{table}

\end{document}